\documentclass{emulateapj}

\shorttitle{Iron-line light echo}
\shortauthors{Komossa et al.}

\begin{document}

\title{Discovery of superstrong, fading, iron line emission and 
       double-peaked Balmer lines of the galaxy SDSSJ095209.56+214313.3  --    
       the light echo of a huge flare}

\author{S. Komossa\altaffilmark{1}, H. Zhou\altaffilmark{1,2}, T. Wang\altaffilmark{2}, 
         M. Ajello\altaffilmark{1}, J. Ge\altaffilmark{3}, J. Greiner\altaffilmark{1},
         H. Lu\altaffilmark{2}, M. Salvato\altaffilmark{4},
         R. Saxton\altaffilmark{5}, H. Shan\altaffilmark{6}, D. Xu\altaffilmark{7}, W. Yuan\altaffilmark{6}}
 
\altaffiltext{1}{Max-Planck-Institut f\"ur extraterrestrische Physik, Postfach 1312, 85741 Garching, Germany; skomossa@mpe.mpg.de}

\altaffiltext{2}{Center for Astrophysics, University of Science and
Technology of China, Hefei, Anhui, 230026, China}

\altaffiltext{3}{Department of Astronomy, University of Florida,
Gainesville, FL 32611}

\altaffiltext{4}{California Institute of Technology, 105-24 Robinson 1200 E. California Blvd., Pasadena, CA 91125} 

\altaffiltext{5}{ESA/ESAC, Apartado 78, 28691 Villanueva de la Canada, Madrid, Spain} 

\altaffiltext{6}{National Astronomical Observatories/Yunnan
Observatory, Chinese Academy of Science, Kunming, P.O. BOX
110, China}

\altaffiltext{7}{National Astronomical Observatories, Chinese Academy of Science,
A20 Datun Road, Chaoyang District, Beijing 100012, China}

\begin{abstract}
We report the discovery of superstrong, fading, high-ionization iron line emission 
in the galaxy SDSSJ095209.56+214313.3 (SDSSJ0952+2143 hereafter), which must have been caused
by an X-ray outburst of large amplitude. 
SDSSJ0952+2143 is unique in its strong multiwavelength variability;
such a broadband emission-line
and continuum response has not been observed before. 
The strong iron line emission 
is accompanied by unusual Balmer
line emission with a broad base, narrow core and double-peaked narrow horns, 
and strong HeII emission. 
These lines, while strong in the SDSS spectrum taken in 
2005, have faded away significantly in 
new spectra
taken in December 2007. 
Comparison of SDSS, 2MASS, GALEX and follow-up GROND photometry reveals 
variability in the NUV, optical and NIR band. 
Taken together, these unusual observations can be explained by a giant outburst
in the EUV--X-ray band, detected even in the optical and NIR.
The intense and variable iron, Helium and Balmer lines represent the ``light echo''
of the flare, as it traveled through circumnuclear material. 
The outburst may have been caused by the tidal disruption of a
star by a supermassive black hole.
Spectroscopic surveys such as SDSS are well suited to detect
emission-line light echoes of such rare flare events.  Reverberation-mapping
of these light echoes can then be used 
as a new and efficient probe of the physical conditions in the 
circumnuclear material in non-active or
active galaxies.     
\end{abstract}


\section{Introduction}

The continuum emission of many active galactic nuclei (AGN) is variable
by a factor of a few. 
The broad-line region (BLR) emission-line response  to changes
in the ionizing continuum provides us with important
information on the size and physical properties of the emission-line
clouds
(see Peterson 2007 for a review).  
A few exceptional cases of non-repetitive, i.e. transient,
extreme broad-line variability have been reported,
including the temporal appearance of an unusually broad HeII line
in NGC\,5548 (Peterson \& Ferland 1986), and of double-peaked Balmer lines
in NGC\,1097 (Storchi-Bergmann et al. 1993) and a few other LINERs;
interpreted as a response to 
transient accretion events. 

Detection of strong continuum and emission-line variability is of great interest,
since it places tight constraints on accretion physics and 
physical conditions in the gas clouds surrounding the supermassive black hole (SMBH). 
High-amplitude variability of emission lines other than
the classical broad lines is rare in AGN. The one notable
exception is the Seyfert galaxy IC\,3599. X-ray bright
during the ROSAT all-sky survey, a near-simultaneous optical
spectrum showed strong high-ionization lines
(Brandt et al. 1995) which were significantly fainter 
in a new optical spectrum taken 8 months later
(Grupe et al. 1995).  
The X-ray flux and the high-ionization
emission lines continued to fade in subsequent years,
and in   
low-state, the galaxy can be classified as
a Seyfert 1.9 or 2. 
The X-ray flare of this active galaxy
was interpreted in terms of a temporarily enhanced accretion rate
(Brandt et al. 1995, Grupe et al. 1995, Komossa \& Bade 1999).  

The highest amplitudes of X-ray variability, up to factors of $\sim$6000, 
have been almost exclusively detected in non-active
galaxies, and these represent the best observational evidence to date for 
the process of tidal disruption of stars by SMBHs 
(e.g., Komossa \& Bade 1999, Halpern et al. 2004,
Komossa et al. 2004). 

Here we report the discovery of very unusual and variable
emission lines of the galaxy SDSSJ0952+2143 at redshift $z$=0.079,
found in a systematic search for emission-line AGN   
in SDSS-DR6 (Sloan Digital Sky Survey Data Release 6; Adelman-McCarthy et al. 2008).  
The optical spectrum of SDSSJ0952+2143 (Fig. 1) looked
very similar, but more extreme, than that of IC3599 in high-state, which is
why we suspected a similar mechanism at work and initiated the
follow-up observations reported in this Letter.
SDSSJ0952+2143 turns out to be unique in 
its broad multi-wavelength continuum
and emission-line response 
to a high-energy outburst. 
We use a cosmology with
$H_{\rm 0}$=70 km\,s$^{-1}$\,Mpc$^{-1}$, $\Omega_{\rm M}$=0.3
and $\Omega_{\rm \Lambda}$=0.7 throughout this Letter.

\section{Multi-wavelength observations and results}

\subsection{SDSS photometry in 2004}

SDSSJ0952+2143 
was brightest during SDSS photometry performed 
on 2004 December 20.
At that time it showed a flat spectral energy distribution (SED) 
with an increase toward
the NIR; inconsistent with the extrapolation of 2MASS data
taken in January 1998 (Fig. 2). 
The SDSS photometric data can be well described by a constant underlying
galaxy SED and a powerlaw component consistent in shape
with that during the SDSS spectroscopy 1 year later, 
but of greater strength.  
The increase towards the NIR can be described
by an extra black body component with a temperature
of 2400 K.

\subsection{SDSS spectroscopy in 2005}

Spectroscopy of SDSSJ0952+2143 was performed on 2005 December 30.
It was fainter than during the photometry.
The continuum SED can be decomposed into a host-galaxy contribution,
and a powerlaw of index ${\alpha}=-1.0$, where $f_\lambda \propto \lambda^{\alpha}$.
The bulk of the variability between Dec. 2004 and Dec. 2005 can be
attributed to a variable powerlaw component which was
fainter by a factor of 2 in 2005.   

During the 2005 spectroscopy, SDSSJ0952+2143  
displayed a rich and unusual emission-line 
spectrum. A variety of high-ionization iron coronal lines 
from [FeVII] and up to [FeXIV] can be identified,
several with unprecedented strengths relative to [OIII]5007, and several 
as strong as [OIII] (Fig. 1 and Tab. 1). 
Transitions include [FeVII]3586,3759, rarely identified in AGN.  
Iron lines consist of a broad
(FWHM $\approx 700-800$ km\,s$^{-1}$) and a narrow (FWHM $\approx 100-150$ km\,s$^{-1}$)
component. 
Strong HeII4686 emission is present
with a width similar to the broad component of the iron lines (FWHM(HeII)=810 km\,s$^{-1}$).
The Balmer line profile shows multiple components. A multi-Gaussian
decomposition reveals 
a broad base (FWHM(H$\alpha_{\rm b}$)=1930 km\,s$^{-1}$,
where the index b denotes the broad component of the profile),
a narrow core (FWHM fixed to that of [SII]6716,31; 200 km\,s$^{-1}$),
and two remarkable unresolved narrow horns (Fig. 1). 
The broad base is redshifted by 570 km\,s$^{-1}$
and has a luminosity of $L_{\rm H\alpha_{\rm b}}=3\,10^{41}$ erg\,s$^{-1}$. 
The two narrow horns are also prominent in H$\beta$,
while its broad  
component is barely detected,
implying a ratio H$\alpha_{\rm b}$/H$\beta_{\rm b}$=9.5. 

Using stellar absorption lines detected in the spectrum,
we have determined the stellar velocity
dispersion, $\sigma_*=95$ km\,s$^{-1}$. 
This implies a black hole mass of 
$M_{\rm BH} = 7\,10^6$ M$_{\odot}$ (Ferrarese \& Ford 2005).

\subsection{GALEX UV photometry in 2006}

The GALEX (Martin et al. 2005) UV photometric data points, of 2006 March 2, 
are above the expected host-galaxy contribution,
and are consistent with the extrapolated powerlaw component
measured during SDSS spectroscopy $\sim$2 months earlier (Fig. 2).

All these results locate the maximum of the outburst 
after the 2MASS observation in 1998,
and possibly close to the 2004 SDSS photometry high-state.

\subsection{Optical and X-ray follow-ups in 2007 and 2008}
In order to search for line variability,
we have taken new optical spectra on 2007 December 4 and 5 
with the OMR spectrograph equipped with a 200\AA/mm grating 
at the Xinglong 2.16m telescope.
In these new optical spectra
most of the iron lines, the HeII line, 
and the broad H$\alpha$ line are significantly fainter
{\em relative to [OIII]} (Fig. 2). 
Broad H$\alpha$, HeII, and [FeXIV] have faded by a factor of $\sim$2. 
[FeX] is no longer detected. Despite 
the lower S/N of the Xinglong spectrum and the coincidence of
telluric atmospheric absorption with the redshifted 
[FeX], this still implies line variability by a least a factor of two.  

New photometry was performed with GROND, a multi-channel imager (Greiner et al. 2008) 
attached to the 2.2m telescope
at La Silla in the SDSS filters g, r, i, and z 
on 2008 January 1, at a seeing of 1$^{\prime\prime}$. 
These data confirm the decline in the optical and NIR band. 
The core of SDSSJ0952+2143 (the central 1.4$^{\prime\prime} \times 1.4^{{\prime\prime}}$)
was a factor 2 weaker than during the 2004 SDSS photometry.  

We have initiated a {\sl Chandra} DDT X-ray observation 
of 10 ks duration which was carried out on 2008 February 5. 
SDSSJ0952+2143 was detected with an ACIS-S countrate of 0.0007 cts/s,
corresponding to a (2--10)\,keV X-ray luminosity of 4\,10$^{40}$ erg/s
(S. Komossa et al., in preparation).  

\section{Discussion}

We have detected the emission-line light echo and the low-energy (NUV, optical,
NIR) SED tail  of a high-energy (EUV, X-ray) outburst of the galaxy
SDSSJ0952+2143. The EUV -- X-ray part of the flare was not observed directly,
but the presence of a strong ionizing continuum at
high-state is implied by the strong 
H$\alpha$ and HeII lines and by the large iron-line fluxes
relative to [OIII]. Further, the ionization potentials of HeII and of
the iron lines (up to 358 eV; FeXIV) imply that the ionizing
continuum extended into the soft X-ray regime. 

While X-ray variability is common in AGN, amplitudes exceeding
a factor of $\sim$30--50 are rare (e.g., Komossa 2002) and they
almost never come with a reported measurable emission-line response.
Outbursts of high amplitude are of special interest, because they
provide tight constraints on accretion physics under extreme conditions.
Furthermore, as the flare travels through the circumnuclear material, its
emission-line light echo puts us in the rare situation to be able to 
perform ``reverberation mapping'' not only of the BLR, but potentially also
of the accretion disk region, the coronal-line region (CLR), the molecular torus, 
and the narrow-line region (NLR)
of active galaxies (Ulmer 1999, Komossa \& Bade 1999), 
and it allows us to detect which of these regions
are also present in non-active galaxies (Komossa 2002).   
 
\subsection{Amplitude of variability, and galaxy classification}

Since  X-ray monitoring simultaneous with and after the SDSS observations
does not exist, we can only indirectly estimate the amplitude of variability
of the outburst.
The fluxes of the iron lines, relative to [OIII], observed in
2005 (1 year after the highest observed state in 2004), are much
higher than those in any other AGN in SDSS-DR6.  
The observed (2005; post-high-state) ratio [FeX]/[OIII] is  
a factor of 60 (20) higher than the Seyfert 2 (Seyfert 1) average
of Nagao et al. (2000; their Tab. 6), implying a high amplitude
of variability. 

It is possible that all emission lines were excited 
by the flare{\footnote{Given
light-travel time delays, the bulk of any permanent low-density gas
at NLR-typical distances will only be reached after a long time. However,
the fraction of this gas that is close to our line
of sight will already be illuminated by the flare.
This mechanism would cause temporary AGN-like line emission even 
in intrinsically
non-active galaxies.
}}.
Alternatively, 
SDSSJ0952+2143 may show permanent low-level activity
traced by low-ionization lines.
The ratio [OIII]/H$\beta_{\rm n}$=2.5
puts the galaxy at the border between AGN and LINER.

\subsection{NIR variability}

While the variable optical continuum likely is the tail of the
high-energy continuum, the variable NIR continuum during the
2004 SDSS photometry high-state is best fit by an
independent 
black body component of $T \simeq 2400$ K, 
perhaps
a signal of dust reprocessing before  destruction. 
The measured temperature slightly exceeds the sublimation
temperature $T_{\rm s}$ of standard silicates and graphite 
(e.g., Granato \& Danese 1994), but non-standard
grain sizes and decomposition may raise the limit on $T_{\rm s}$.

\subsection{Iron line diagnostics}

For the first time, transitions of [FeVII]3586,3759 as strong as [OIII] have been
identified in a galaxy spectrum.  
Several [FeVII] line ratios can be used to estimate gas temperature and gas density. 
The observed high-state ratio of [FeVII]5158/[FeVII]6087 $\approx$ 0.2 
implies a density of $\log n_{\rm e} = 6-7$.  
The ratio [FeVII]3759/[FeVII]6087=1 implies a temperature
of at least $T \ge 20000$ K,  
and up to $T = 50000$ K if $\log n_{\rm e} \le 6$  (Keenan \& Norrington 1987). 
The high density and temperature inferred and the variability of the iron lines on the timescale
of years implies an emission-region close to the active nucleus.
Possible sites of origin of the observed lines are the
inner wall of the dusty torus, and/or the tidal
debris of a disrupted star flung out from
the system, and interacting with its environment (Khoklov \& Melia 1996). 
Monitoring of the fading lines will provide a key diagnostic of the
physical conditions in the line-emitting region;
if this region corresponds to the dusty torus, we can do ``torus
reverberation mapping'' this way.

\subsection{Double-peaked Balmer line emission}

The H$\alpha$ and H$\beta$ lines show a complex profile which
can be decomposed into at least four components: 
a broad component
with a width of 1930 km\,s$^{-1}$ 
redshifted by an extra 570 km\,s$^{-1}$, 
a narrow component at the same redshift as other narrow forbidden lines,  
and two narrow horns, the more conspicuous one redshifted by 540 km\,s$^{-1}$. 
Double-peaked Balmer lines have been detected in a number of AGN (see
Eracleous et al. 2006 for a review). However, generally, it is the {\em broad} component
of the Balmer lines which is double peaked.
The strong narrow horn has no counterpart in any other emission line; it is only
detected in H$\alpha$ and H$\beta$. One possible explanation is that it arises
in a region of very high density (such as the stellar post-disruption debris, 
or an accretion disk; see Sect. 3.5) 
where forbidden lines are suppressed. The summed redshifted
broad and narrow component of the H$\alpha$ profile resembles to some extent,
but not in detail, relativistic FeK$\alpha$ line profiles seen in some
AGN X-ray spectra (review by Miller 2007). 
The broad Balmer decrement, H$\alpha$/H$\beta$=9.5, likely implies 
a significant contribution from collisional excitation indicative of
high density, or else is affected by optical depth effects.

\subsection{Outburst mechanism}

The highest amplitudes of variability detected to date, up to a factor of $\sim$6000,
have essentially all been observed from non-active galaxies, and have been
interpreted in terms of tidal disruptions of stars by supermassive black holes
(Komossa \& Bade 1999, Halpern et al. 2004, Komossa et al. 2004, and references therein).
Stars approaching a SMBH will be tidally disrupted once the tidal
forces of the SMBH exceed the star's self-gravity,
and part of the stellar debris will be accreted, producing a luminous flare
of radiation which lasts on the timescale of
months to years (e.g., Carter \& Luminet 1982,
Rees 1988;    
see Komossa 2002 for a much longer list of references on theoretical aspects 
of tidal disruption). 

The outburst mechanism of SDSSJ0952+2143 
is either related to processes in the accretion disk such as 
an instability
(only possible if SDSSJ0952+2143 is permanently active), or is else related to 
the tidal disruption of a star by the SMBH at its center (possible in both cases,
active or non-active galaxy); see Komossa \& Bade (1999) for a discussion
of relevant timescales of both mechanisms. 
At present, we cannot tell whether SDSSJ0952+2143 harbors
a permanent low-luminosity AGN, traced by [OII] and [OIII], or whether these
lines were also excited by the flare and represent 
low-density ISM/NLR emission close to our
line of sight.

In the context of tidal disruption, 
the unusual profile and double-peakedness of the Balmer lines can
be understood 
as emission from the accretion disk formed from
the stellar debris after disruption.    
While Balmer line emission from 
different streams of the
disrupted star looks complex during the first several months
after disruption (Bogdanovic et al. 2004),
it would appear more like a normal disk-line thereafter.  

While automized procedures are currently being set up to identify 
new flares rapidly in the course of current  
X-ray surveys
(Yuan et al. 2006, Esquej et al. 2007),   
spectroscopic surveys such as SDSS-II will allow us to
perform independent flare searches by the detection 
of variable emission-line signatures. 
SDSSJ0952+2143 is exceptional among the few known flaring galaxies
in its strong multi-wavelength variability, 
encompassing the NIR, optical 
and NUV continuum SED, allowed and forbidden optical and NIR emission lines,
and the EUV and X-ray SED.
Such a broadband line and continuum response has not been 
detected before.
Multi-wavelength monitoring of objects such as SDSSJ0952+2143 will enable us to
follow the fading of the emission lines, providing a unique light echo
mapping of the circumnuclear material potentially including the accretion disk, BLR, torus, and ISM;
and flung-out post-disruption stellar material if the cause of the flare was stellar tidal disruption. 

%


\begin{deluxetable}{ccccccccccc} 
\tabletypesize{\tiny}
\tablecaption{Emission-line ratios relative to [OIII]5007}
\tablewidth{0pt}
\tablehead{
\colhead{[FeVII]3759} & {HeII4686} & {H$\beta_{\rm totl}$} & {H$\beta_{\rm n}$} & {[OIII]} & {[FeXIV]5303} & {[FeVII]6087} &
{[FeX]6375} & {H$\alpha$} & {[SII]6725}\tablenotemark{a} & {[FeXI]7892} } 
\startdata
0.7 & 0.7 & 1.3\tablenotemark{b}& 0.4 & 1.0 & 0.4 & 0.7 & 1.3 & 10.0\tablenotemark{b} & 0.5 & 1.0 
\enddata
\tablenotetext{a}{Sum of [SII]6716 and [SII]6731.}
\tablenotetext{b}{Including the broad and narrow component, but not the two extra horns.}
\end{deluxetable}


\begin{figure*}[b]
\plottwo{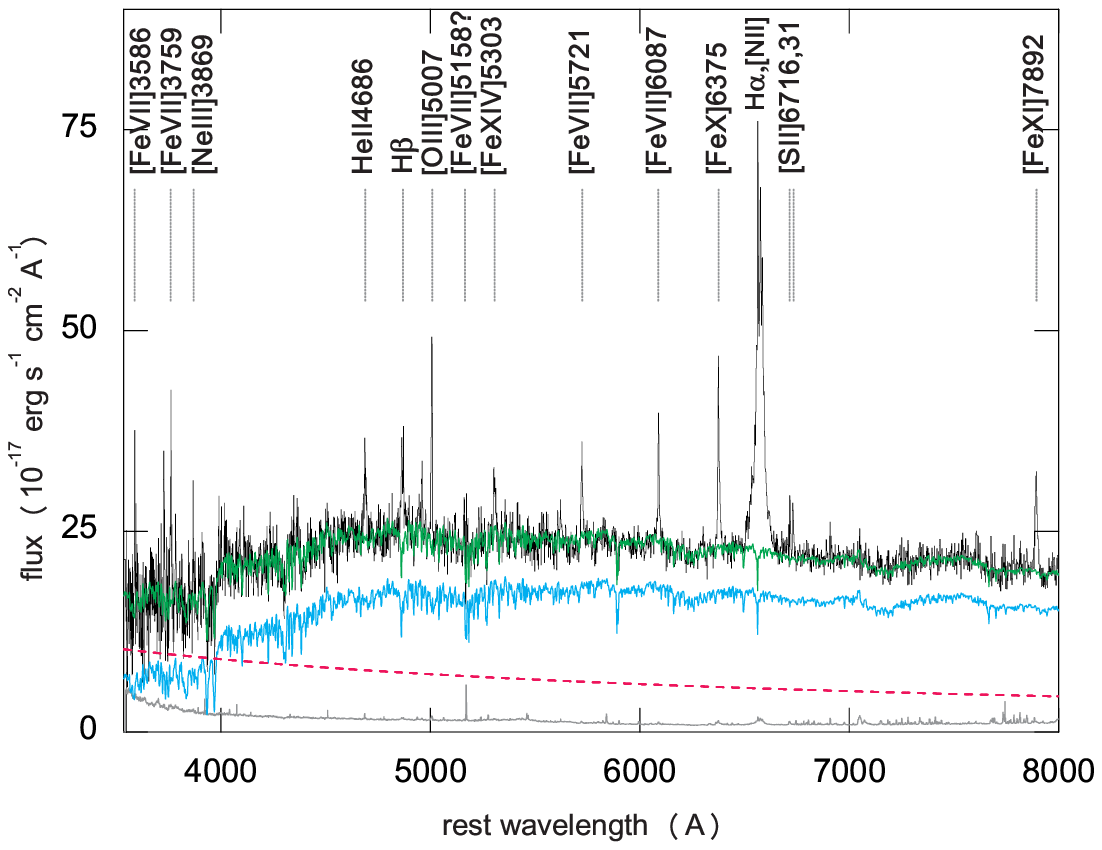}{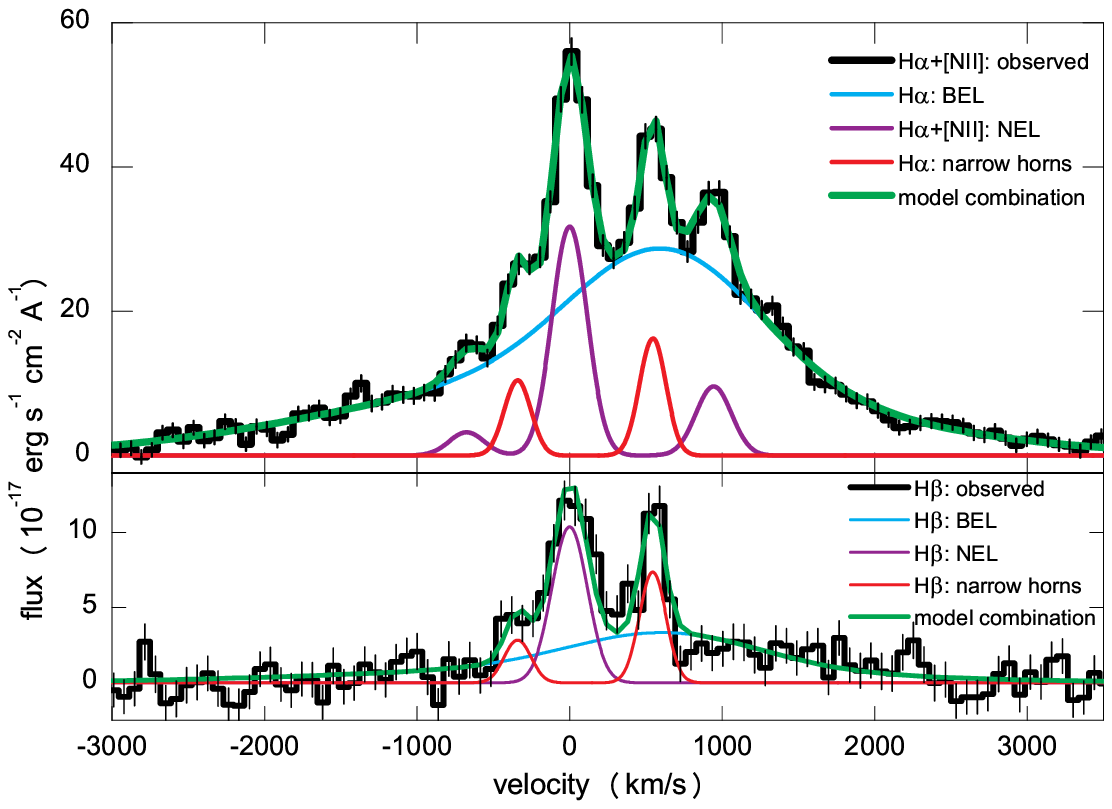}
\caption{Left: SDSS spectrum of SDSSJ0952+2143 (black),  
and continuum decomposition into stellar (blue) and powerlaw component (red).
Right: profiles of H$\alpha$ and H$\beta$, and 
decomposition into broad base (BEL), narrow core (NEL), and two narrow horns.
}
\end{figure*}

\begin{figure*}
\plottwo{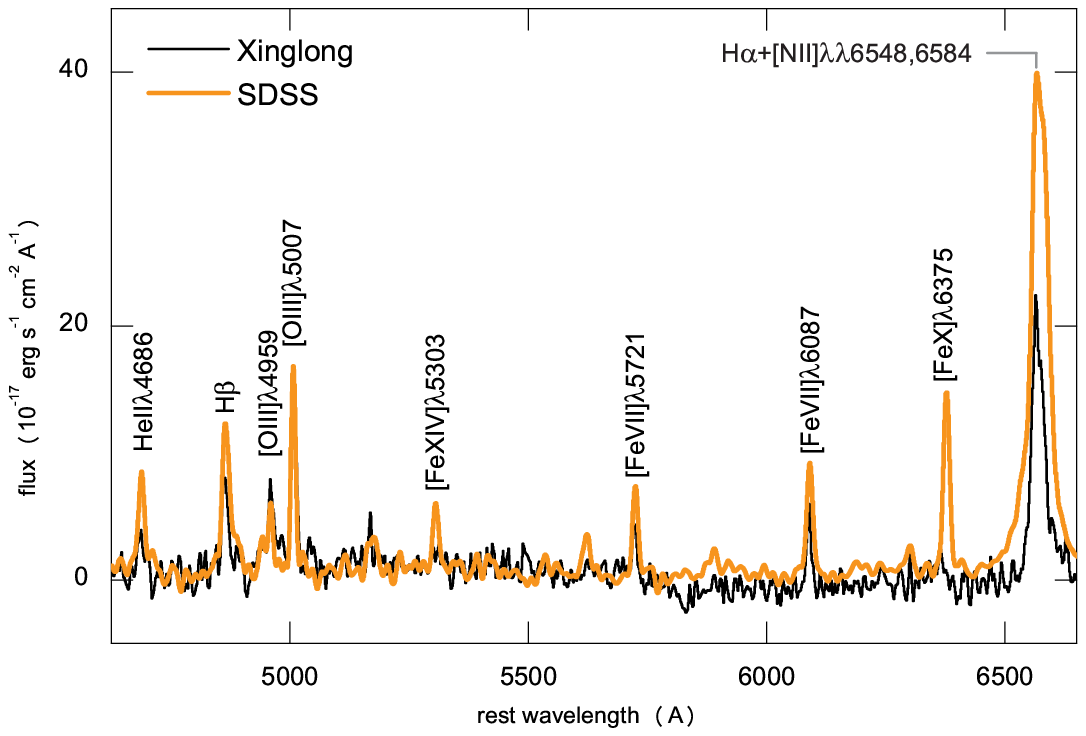}{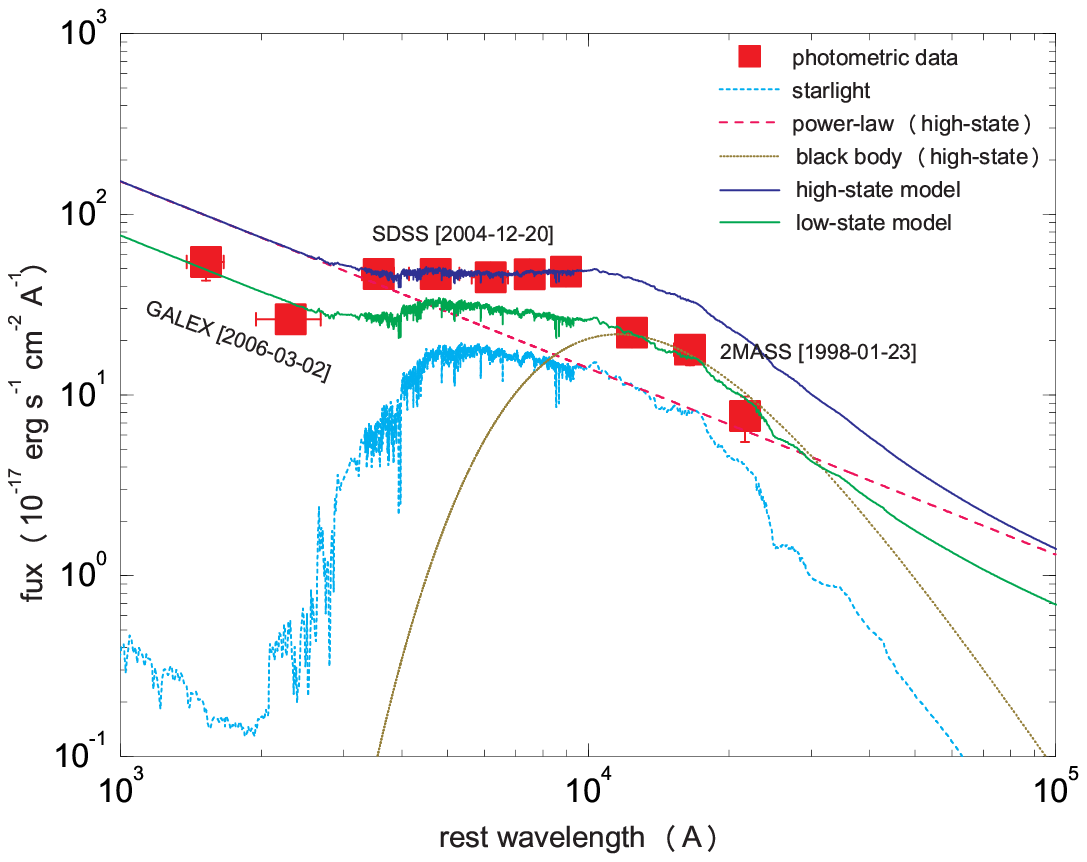} 
\caption{Left: Comparison of the new Xinglong spectrum of Dec. 2007 (black) 
with the previous SDSS spectrum (orange; the resolution was 
degraded to match that of Xinglong). Both spectra have been continuum-subtracted,
and scaled assuming constant [OIII] flux. 
Right: UV-optical-NIR SED; data points and model fits as marked in the plot, 
taken at different epochs. 
}
\end{figure*}

\end{document}